\documentclass[fleqn,usenatbib]{mnras}

\usepackage[T1]{fontenc}

\DeclareRobustCommand{\VAN}[3]{#2}
\let\VANthebibliography\thebibliography
\def\thebibliography{\DeclareRobustCommand{\VAN}[3]{##3}\VANthebibliography}

\usepackage{graphicx}	
\usepackage{amsmath}	
\usepackage{amssymb}	

\renewcommand{\P}{\mathcal{P}}

\usepackage{color}
\usepackage[caption=false]{subfig}
\def\beq{\begin{eqnarray}}
\def\eeq{\end{eqnarray}}

\definecolor{darkgreen}{RGB}{0,120,0}
\newcommand{\new}[1]{#1}
\newcommand{\resub}[1]{#1}

\usepackage{multirow}
\usepackage{booktabs}

\title[An Exact Solution of Kepler's Equation]{Kepler's Goat Herd: An Exact Solution \resub{to Kepler's Equation for Elliptical Orbits}}

\author[O.H.E. Philcox \textit{et al.}]{Oliver H.\,E. Philcox$^{1,2}$\thanks{E-mail: \href{mailto:ohep2@cantab.ac.uk}{ohep2@cantab.ac.uk} (OP)},
Jeremy Goodman$^{1}$ 
and Zachary Slepian$^{3,4}$
\\
$^{1}$Department of Astrophysical Sciences, Princeton University, Princeton, NJ 08540, USA\\
$^{2}$School of Natural Sciences, Institute for Advanced Study, 1 Einstein Drive, Princeton, NJ 08540, USA\\
$^{3}$Department of Astronomy, University of Florida, 211 Bryant Space Science Center, Gainesville, FL 32611, USA\\
$^{4}$Physics Division, Lawrence Berkeley National Laboratory, 1 Cyclotron Road, Berkeley, CA 94709, USA\\
}

\pubyear{2021}

\usepackage{newtxtext,newtxmath}
\begin{document}
\label{firstpage}
\pagerange{\pageref{firstpage}--\pageref{lastpage}}
\maketitle
\begin{abstract}
A fundamental relation \new{in} celestial mechanics is Kepler's equation, linking an orbit's mean anomaly to its eccentric anomaly and eccentricity. Being transcendental, \new{the equation} cannot be directly solved for \new{eccentric anomaly} by conventional treatments; \new{much work has been devoted to} approximate methods. Here, we give an explicit integral solution, utilizing methods recently applied to the ``geometric goat problem'' and \new{to} the dynamics of spherical collapse. The solution is given as a ratio of contour integrals; these can be efficiently computed via numerical integration for arbitrary eccentricities. The method is found to be highly accurate in practice, with our \new{\textsc{C++}} implementation outperforming conventional root-finding and series approaches by a factor \new{greater than two}.
\end{abstract}




\section{Introduction}\label{sec: intro}
Kepler's equation states that
\beq\label{eq: Kepler-equation}
    E - e \sin E = \ell
\eeq
for eccentric anomaly $E\in[0,2\pi)$, \new{eccentricity $e\in[0,\infty)$} and mean anomaly $\ell\in[0,2\pi)$ (often denoted $M$). This describes the dynamics of a two-body system, and \new{was first published} over 400 years ago \citep{kepler1609}. It is most commonly applied to (non-circular) elliptical orbits, whereupon $0<e<1$. Assuming such an orbit, the quantities in \eqref{eq: Kepler-equation} may be related to other orbital parameters of interest via
\beq\label{eq: cartesian}
    \ell = n(t-\tau), \quad  x = a(\cos E-e), \quad y = a\sqrt{1-e^2}\sin E
\eeq
\citep{1999ssd..book.....M}, \new{with $t-\tau$ being} the time-coordinate relative to pericenter crossing, $n$ the mean motion, $a$ the semi-major axis, and $(x,y)$ the Cartesian positions in the orbital plane, relative to the center of the ellipse. A solution of \eqref{eq: Kepler-equation} thus specifies the full location of the orbiting body as a function of time.

Despite its age, Kepler's equation is no less relevant today than in 1609; it still appears in many contexts, with crucial examples \resub{including Global Positioning System (GPS) calibration and} the determination of satellite and debris positions by the \new{United States Space Surveillance Network (SSN)\footnote{See \href{https://www.space-track.org/}{www.space-track.org}.}}. Being a transcendental equation, conventional methods cannot solve \eqref{eq: Kepler-equation} directly, which has led to a vast literature concerning alternative approaches, such as root-finding and series solutions. Indeed, \citet{1993sket.book.....C} notes that new methods of solution have \new{been} proposed in almost every decade since 1650. Below, we present our contribution to this canon.

Before discussing Kepler's equation in detail, we briefly mention a seemingly unrelated topic: the `geometric goat problem'. Roughly speaking, the problem is the following: "Imagine a goat is tied to the edge of a circular enclosure. How long must the rope be such that the goat can graze in exactly half of the field?" Recent work by \citet{Ullisch} shows this to be equivalent to solving the transcendental equation $\sin\beta - \beta\cos\beta = \pi/2$ for $\beta$. Using methods drawn from complex analysis, \citet{Ullisch} provides an explicit solution to the problem via a ratio of contour integrals, each of which can be easily computed using FFTs or direct summation. Given the similarity between this and \eqref{eq: Kepler-equation}, we may ask whether an analogous solution is possible in our case. Further \new{support for this approach is given by} \citet{2021arXiv210309823S}, which used the same methods to explicitly solve the spherical collapse equations, \new{which are a special case of \eqref{eq: Kepler-equation} with $e = 1$.}

\vskip 4pt
The remainder of this paper is structured as follows. In \S\ref{sec: integral-solution} we present the explicit solution of Kepler's equation, alongside a proof of its validity, before discussing its practical evaluation in \S\ref{sec: practical}. Discussion of alternative solution methods is provided in \S\ref{sec: comparison} before a comparison is made in \S\ref{sec: discussion} alongside our conclusions. All computations in this work were performed in \new{\textsc{C++}} and plotted using \textsc{python}.

\section{Integral Solution}\label{sec: integral-solution}
Our first step is to demonstrate that \eqref{eq: Kepler-equation} has a unique solution in the range of interest. We will assume that the eccentricity is both fixed and in the range $e\in(0,1)$, \textit{i.e.} that we have a non-circular elliptical orbit. 
Considering the function
\beq\label{eq: gE-def}
    g(E;e)\equiv E - e \sin E
\eeq
the solution $E^*(\ell,e)$ satisfies $g(E_*(\ell,e);e) = \ell$. \resub{Given that $g(E;e)$ is an odd function of $e$, and \eqref{eq: Kepler-equation} is invariant under $E\to E+2\pi$, $\ell\to\ell+2\pi$, we need only consider the range $\ell\in[0,\pi]$. We additionally avoid the trivial solutions $E = 0, \pi$ at $\ell = 0, \pi$.}

\begin{figure}
    \centering
    \includegraphics[width=0.55\textwidth]{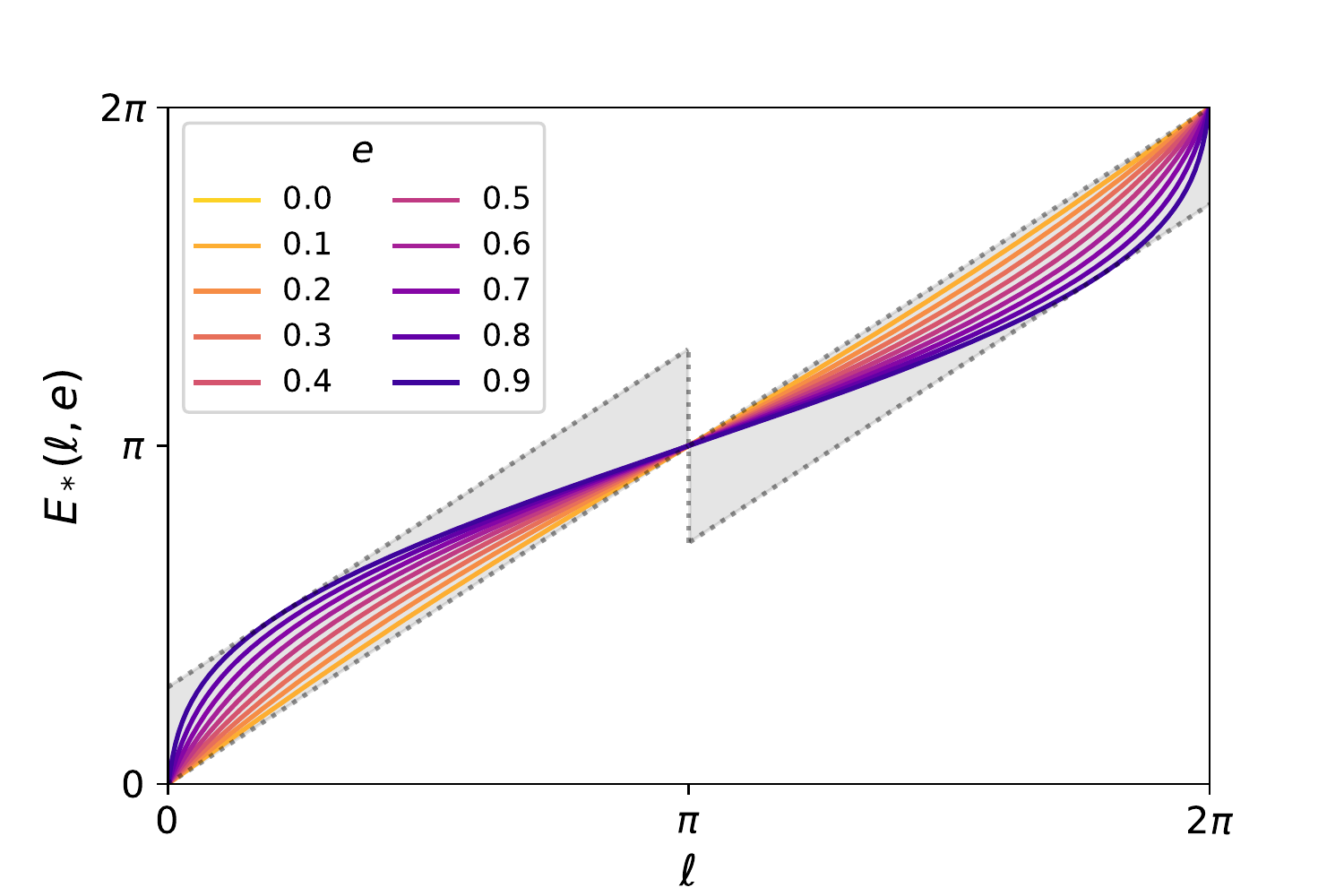}
    \caption{Solutions to Kepler's equation as a function of the mean anomaly $\ell$ for various values of the eccentricity $e$ (shown in the legend). All results are obtained using the contour integration method of this work with 32 steps. Dotted lines show the functions \new{$E = \ell$} and $E = \ell\pm e$; the solution for $E$ lies \new{in the light gray region between the two} in all cases. \resub{As noted in \S\ref{sec: integral-solution}, the solutions are symmetric about $\ell = \pi$.}}
    \label{fig: solutions}
\end{figure}

Considering the end-points, $g(0;e) = 0$, \resub{$g(\pi;e) = \pi$}; since the function is smooth, $g(E;\ell,e)$ must thus equal $\ell$ at least once for all $\ell \in (0,\pi)$. In fact $g(E;e)$ is monotonic in $E$ for all $|e|<1$, hence there is a single solution $E_*(\ell,e)$. This is easily shown by noting that both $E$ and $e\sin E$ are zero at $E = 0$ and $E$ increases faster than $|e\sin E|$ for all $|e|<1$, \new{hence $E>|e\sin E|$ everywhere}. For reference, we provide a plot of the solution $E_*(\ell,e)$ for various values of $e$ and $\ell$ in Fig.\,\ref{fig: solutions}.

To define an integral solution to \eqref{eq: Kepler-equation}, we first consider the entire function $f$
\beq\label{eq: f-z-def}
    f&:&U\rightarrow\mathbb{C}\\\nonumber
    f(z; \ell,e) &=& z - e \sin z - \ell 
\eeq
defined on a simply-connected region $U=(0,2\pi)\times i(-\infty,\infty)$ in the complex plane. The parameters $e,\ell$ take the values $e\in(0,1)$ and \resub{$\ell\in(0,\pi)$} as before. Restricting to the real line, the solution to $f(z; \ell,e)=0$, denoted $z_0(\ell,e)$, recovers Kepler's equation, in the sense that $z_0(\ell,e) \equiv E_*(\ell,e)$.

To proceed, we use the theorem of \citet{Ullisch}, as in \citet{2021arXiv210309823S} (and found previously in \citealt{Jackson16,Jackson17,2002SIAMR..44..227L,Luck15}).\footnote{\resub{An alternative solution using complex methods can be found in \citet{1972CeMec...6..294S}. Our approach is significantly more straightforward however.}} This states that, for an open simply connected subset $U$ of $\mathbb{C}$ with \new{a} non-zero analytic function $f$ defined therein, for each simple zero $z_0\in U$ of $f$, there exists a closed curve $C\in U$ such that:
\beq\label{eq: z0-def}
    z_0 = \left.\left[\oint_C\frac{z\,dz}{f(z)}\right]\right/\left[\oint_C\frac{dz}{f(z)}\right].
\eeq
\new{\eqref{eq: z0-def} follows directly from the Residue Theorem under the assumptions stated: the numerator evaluates to $2\pi i z_0/f'(z_0)$, and the denominator to $2\pi i/f’(z_0)$.} Furthermore, \eqref{eq: z0-def} holds for \textit{any} Jordan curve $C\in U$, provided that (a) it encloses $z_0$, and (b) $f(z)\neq 0$ at all points on $C$ and its interior, except at $z=z_0$.

In our context, $f(z)\equiv f(z,\ell,e)$ is defined by \eqref{eq: f-z-def}, and, provided certain conditions are met, we may use \eqref{eq: z0-def} to evaluate $z_0(\ell,e)$ and hence obtain $E_*(\ell,z)$. Below, we consider the conditions in detail:
\begin{enumerate}
    \item \textbf{$z_0$ lies within $U$}. By the discussion above, $f(z,\ell,e)=0$ on the real-line at a single point with $z_0(\ell,e)=E_*(\ell,e)$. Since \resub{$E\in(0,\pi)$} (for \resub{$\ell\in(0,\pi)$}), \new{$z_0$} lies within $U$. 
    
    \item \textbf{$z_0$ is a simple zero}. From \eqref{eq: f-z-def}, $f'(z) = 1-e\cos z$, which, for $e\in(0,1)$, has no solutions on the real line. Thus $f'(z_0)\neq 0$, hence the zero at $z_0$ is simple.
    
    \item \textbf{$z_0$ is enclosed within $C$}. A simple choice is to assume the (Jordan) circular contour \resub{$C=\{z:|z-\pi/2|<\pi/2-\epsilon\}$} for small parameter $\epsilon>0$ (analogous to \citealt{2021arXiv210309823S}). This intersects the real axis at \resub{$\mathrm{Re}[z] = \epsilon,\pi-\epsilon$}; thus, for sufficiently small $\epsilon$, $z_0$ is enclosed with $C$ (for any choice of $e$ and $\ell$). In practice, \new{we find a different contour to be more advantageous for efficient numerical implementation; this is} discussed below.
    
    \item \textbf{$z_0$ is the \textit{only} zero within $C$}. To prove this, we first consider the region $\mathcal{R}\subset U$ defined by $\mathcal{R}=(0,2\pi)\times i(-M,M)$ for large $M$. \new{Since $C$ lies within $\mathcal{R}$}, \new{the condition is satisfied if} $f(z;\ell,e)\neq 0$ everywhere in $\mathcal{R}\backslash\{z_0\}$.\footnote{The notation $\mathcal{R}\backslash\{z_0\}$ indicates the set $\mathcal{R}$ excluding the point $z_0$.} To prove \new{this}, we consider the variation in $\mathrm{arg}[f(z;\ell,e)]$ as one traverses the boundary $\partial\mathcal{R}$ in a counter-clockwise fashion. Writing $z = (x+iy)$, \new{$f(z;\ell,e)$} may be written \new{$f(x,y;\ell,e) = \left(x-e\sin x\cosh y-\ell\right)+i\left(y-e\cos x\sinh y\right)$}. Considering \new{each part of the contour} in turn, \new{and taking $M\to\infty$}:
    \begin{itemize}
        \item Moving from $z=0-iM$ to $z=2\pi-iM$, we have \new{$f(z;\ell,e)= (i/2)e\exp(M+ix)+\mathcal{O}(M)$}, thus $\Delta \mathrm{arg}\left[f(z;\ell,e)\right]=+2\pi$.
        \item From $z=2\pi-iM$ to $2\pi+iM$, \new{$f(z;\ell,e)= 2\pi-\ell+i(y-e\sinh y)$}, thus $\mathrm{Re}[f(z;\ell,e)]$ is constant and positive, whilst $\mathrm{Im}[f(z;\ell,e)]$ moves from \new{$(1/2)e\exp{M}$ to $-(1/2)e\exp M$}, giving $\Delta \mathrm{arg}\left[f(z;\ell,e)\right] = -\pi$. 
        \item Similarly from $z=2\pi+iM$ to $0+iM$, \new{$f(z;\ell,e)= -(i/2)e\exp(M-ix)+\mathcal{O}(M)$}, thus $\Delta\mathrm{arg}[f(z;\ell,e)]=+2\pi$.
        \item Finally, from $z=0+iM$ to $z = 0-iM$, \new{$f(z;\ell,e)= -\ell+i(y-e\sinh y)$}, thus $\mathrm{Re}[f(z;\ell,e)]$ is constant and negative, whilst \new{$\mathrm{Im}[f(z;\ell,e)]$ moves from $-(1/2)e \exp M$ to $(1/2)e\exp M$}, giving $\Delta \mathrm{arg}\left[f(z;\ell,e)\right]=-\pi$.
    \end{itemize}
    Summing the contributions, we find $\Delta\mathrm{arg}[f(z;\ell,e)]=+2\pi$ when traversing the full closed contour $\partial \mathcal{R}$ in a counter-clockwise fashion. From the argument principle, this is equal to $2\pi(Z-P)$ where $Z$ and $P$ are \new{respectively} the number of zeros and poles within $\mathcal{R}$. Since $P=0$ (as neither $z$ nor $\sin z$ contain any poles in $\mathbb{C}$), we find $Z = 1$, thus $\mathcal{R}$ contains only a single zero. \new{This implies that $z_0$ is the only zero of $f(z;\ell,e)$ in $\mathcal{R}$, or,} given that $M$ is arbitrary, $f(z;\ell,e)\neq 0$ for all $z\in U\backslash\{z_0\}$.
\end{enumerate}

It remains to choose the contour $C$. Whilst the large circle $C = \{z:|z-\pi/2|<\pi/2-\epsilon\}$ is a valid choice, given that it lies within $U$ and encloses $z_0$, for numerical efficiency it is preferred to use \resub{a smaller circular contour} \new{(cf.\,\S\ref{sec: practical})}. To form this, we first consider \new{the function $g(E;e) = E-e\sin E$} (cf.\,\ref{eq: gE-def}) in more detail. In particular, we note that \resub{for $E\in(0,\pi)$:}
\beq
    0<\sin E<1 \quad \Rightarrow \quad E-e<g(E;e)<E
\eeq
\new{Given that our solution requires $g(E^*(\ell,e);e) = \ell$, we find the bound} $E_*(\ell,e)\in(\ell,\ell+e)$ for $\ell<\pi$, as shown in Fig.\,\ref{fig: solutions}. Together, this motivates the contour \resub{$C = \{z:|z-(\ell+e/2)| = e/2\}$}, \textit{i.e.} a circle of radius $e/2$ centered at $\ell+ e/2$. This is guaranteed to enclose $z_0$ (ignoring the trivial case $\ell = \pi$), and is contained within $U$, since $e<\pi/2$.\footnote{\resub{An alternative bound is given by $E_*(\ell,e)\in(\ell,\mathrm{min}[\pi,\ell+e])$. Whilst this results in a tighter contour for $\ell>\pi-e$, it will not be applied in this work, since the associated circular contour has a radius dependent on $\ell$, which hampers numerical efficiency (cf.\,\S\ref{sec: practical}), since the contour integrand is no longer separable in $\ell$ and the path-variable $x$. For further discussion on solution bounds see \citet{1986CeMec..38..111S,1998CeMDA..70..131S}.}}
 
\begin{figure}
    \centering
    \includegraphics[width=0.48\textwidth]{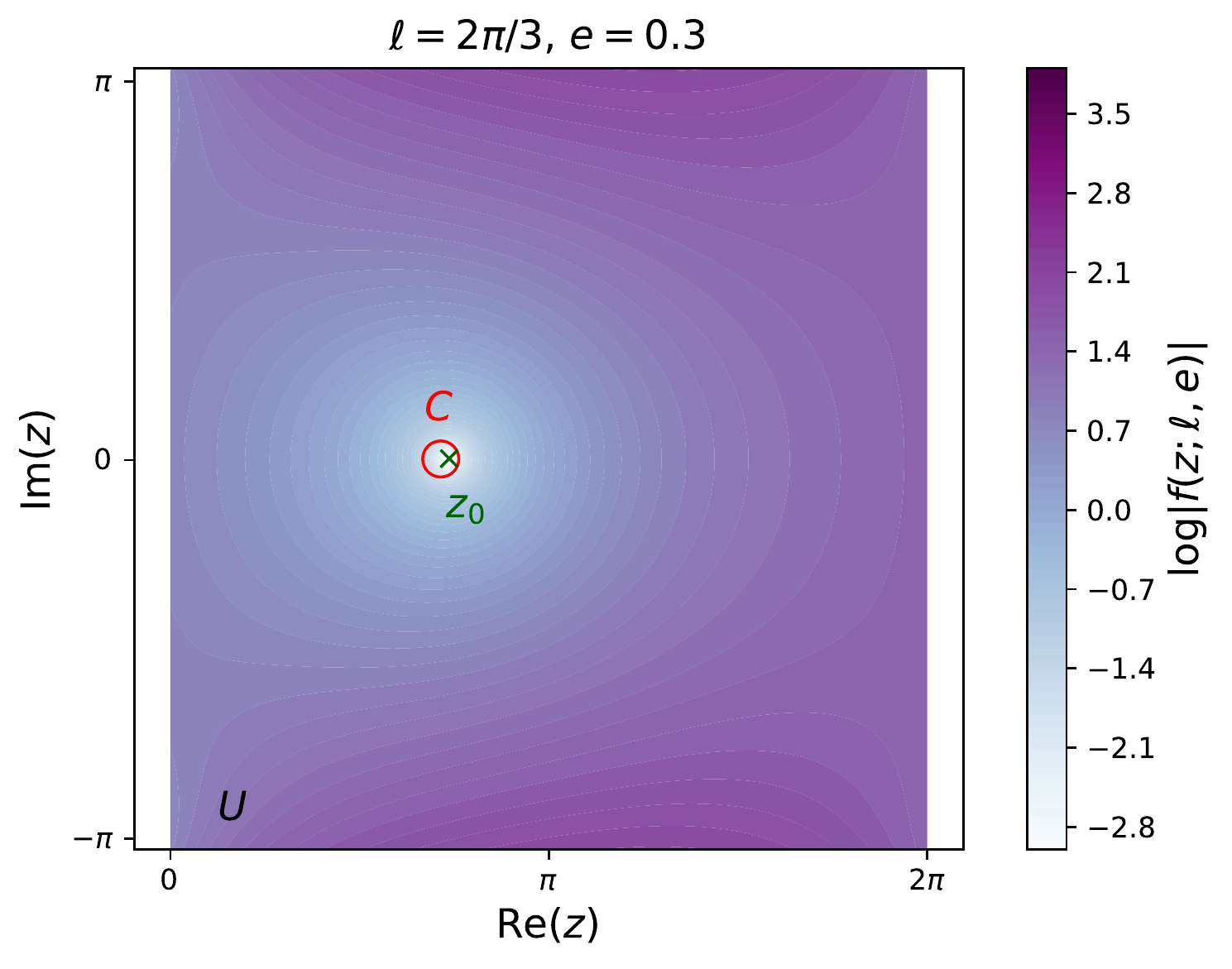}
    \includegraphics[width=0.48\textwidth]{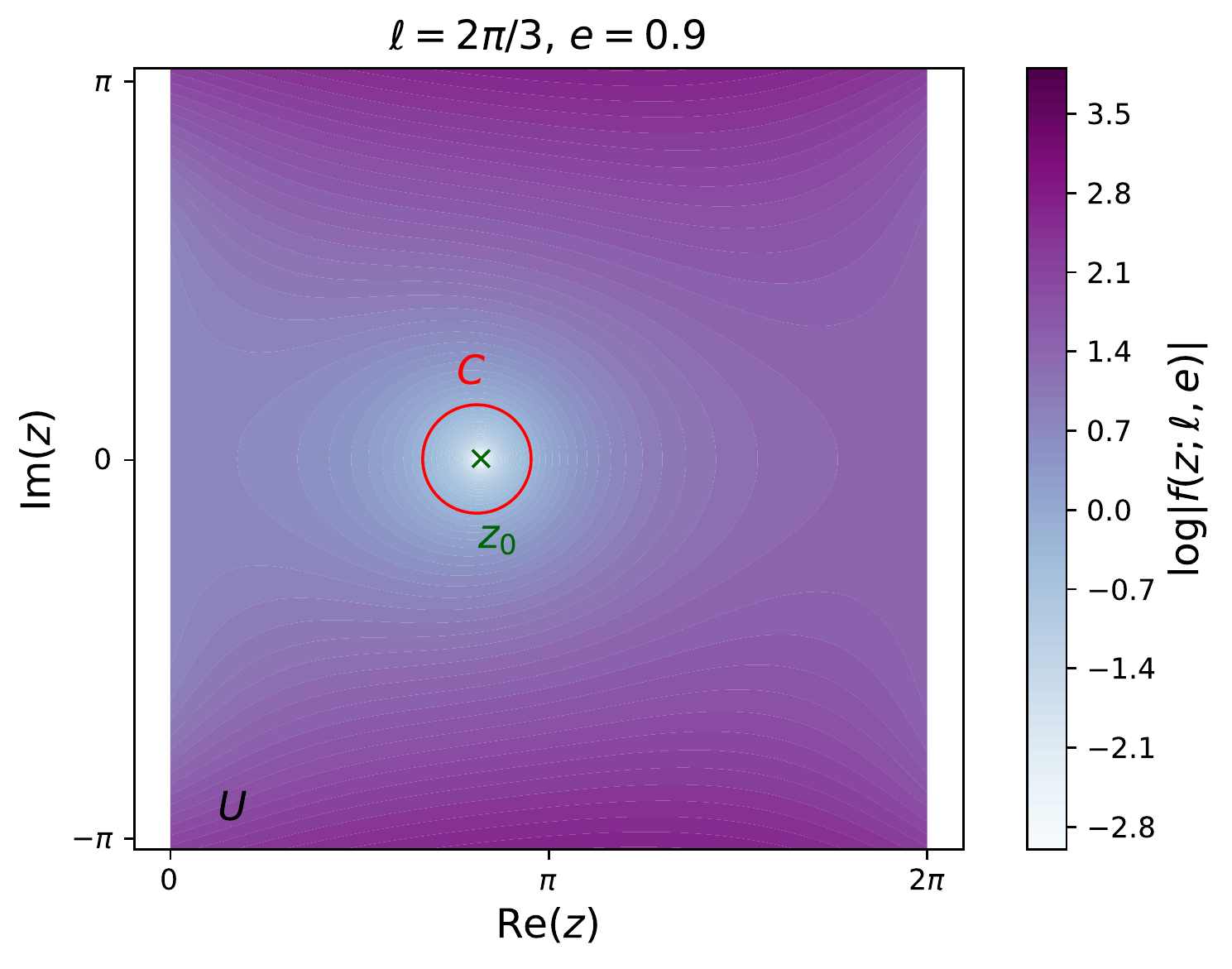}
    \caption{Plots of $f(z;\ell,e)$ (equation \ref{eq: f-z-def}) in the complex plane, from which the integral solution \eqref{eq: integral-solution} to Kepler's equation is obtained. The colorbar shows the value of $\log |f(z;\ell,e)|$ with the green $\times$ indicating $z_0$ such that $f(z_0; \ell,e) = 0$, \textit{i.e.} the desired solution. The colored area indicates the open subset $U$ upon which the result \eqref{eq: z0-def} is valid (denoted by the black $U$ at lower left), and we evaluate the contour integrals on the closed curve $C$ (in red), which is guaranteed to contain $z_0$. As proven in the text, $z_0$ is the sole zero of $f(z;
    \ell,e)$ in $U$. \new{Note that the $C$ encloses the region $E\in(\ell,\ell+e)$ on the real line in which the true solution certainly lies; its center does not have to align with $z_0$ however.} Here, we show the behavior \resub{at $\ell = 2\pi/3$} for two choices of eccentricity ($e = 0.3$ and $e = 0.9$), but all choices are qualitatively similar.}
    \label{fig: contour}
\end{figure}

Given the above, we may apply \eqref{eq: z0-def} to our problem, obtaining the solution
\beq\label{eq: integral-solution}
    E_*(\ell,e) &\equiv& z_0(\ell,e) = \left.\left[\oint_C\frac{z\,dz}{z-e\sin z-\ell}\right]\right/\left[\oint_C\frac{dz}{z-e\sin z-\ell}\right].
\eeq
\resub{This may be straightforwardly inserted into \eqref{eq: cartesian} to obtain the Cartesian evolution of the orbit as a function of $\ell$.} For reference, we show the function $f(z;\ell,e)$ alongside the region $U$ and the contour $C$ in Fig.\,\ref{fig: contour}, for a representative choice of $\ell$ and $e$.

\section{Practical Evaluation}\label{sec: practical}
To obtain a simply implementable form of \eqref{eq: integral-solution}, we first parametrize the circular contour $C$ by the function $\gamma:[0,1)\rightarrow C$ where $\gamma(x) = x_0+\Delta x\,e^{2\pi ix}$ for contour center $x_0$ and radius $\Delta x$. \new{As in \S\ref{sec: integral-solution}}, these are set to \resub{$x_0 = \ell+e/2$, $\Delta x = e/2$}. \new{Inserting this into \eqref{eq: integral-solution} and changing variables leads to}
\beq\label{eq: fft-solution-I}
    E_\ast(\ell,e) &=& x_0 + \Delta x\left.\left[\int_0^1dx\,e^{4\pi ix}a(x;\ell,e)\,dx\right]\right/\left[\int_0^1 dx\,e^{2\pi ix}a(x;\ell,e)\right]
\eeq
where we define
\beq\label{eq: ax-def}
    a(x;\ell,e) = 1/f(\gamma(x);\ell,e) &=& 1/\left[x_0+\Delta x\,e^{2\pi ix}-\ell+e\sin\left(x_0+\Delta x\,e^{2\pi i x}\right)\right].
\eeq
Defining the Fourier series coefficients of \new{$a(x;\ell,e)$} by
\beq\label{eq: ak-def}
    a_{k}(\ell,e) &\equiv& \int_{0}^1 dx\,e^{-2\pi i k x}a(x;\ell,e),
\eeq
for integer $k$, \eqref{eq: fft-solution-I} can be written
\beq\label{eq: fft-solution-II}
    E_\ast(\ell,e) &=& x_0 + \Delta x\,\frac{a_{-2}(\ell,e)}{a_{-1}(\ell,e)}.
\eeq

For any given $\ell$ or $e$, we can thus evaluate $E_\ast(\ell,e)$ by computing the integrals $a_{-2}$ and $a_{-1}$. In general, these must be evaluated numerically. One option for this (advocated for in \citealt{Ullisch} and \citealt{2021arXiv210309823S}) is to use a Fast Fourier Transform (FFT) to obtain all $a_{k}$ frequencies simultaneously, for given $\ell,e$. However, this is \new{not the most efficient choice}. The FFT algorithm scales as $N\log N$, for $N$ grid-points; for computing $N$ frequencies, this is clearly superior to computing $N$ numerical integrals, but when only two are required (as here), the latter method is preferred. To expedite computation, we note that the real-part of each integral is symmetric around $x = 0.5$ (with the imaginary part antisymmetric, and thus cancelling), giving
\beq\label{eq: a-k-approx}
    a_k(\ell,e) &=& 2\int_0^{0.5}dx\,\mathrm{Re}\left[e^{-2\pi ik x}a(x;\ell,e)\right]\\\nonumber
    &\approx& A\left(a(0;\ell,e) + (-1)^k a(0.5;\ell,e) + \sum_{j=1}^{N-1} \mathrm{Re}\left[e^{-2\pi i k j}a(0.5j/N;\ell,e)\right]\right)
\eeq
where we have approximated the integral by a discrete sum of $N$ points in the second line, for some normalization constant $A$ \new{which will cancel in \eqref{eq: fft-solution-II} when we take the ratio}. Furthermore, we may compute the real parts explicitly and thus avoid any invoking complex numbers in the implementation of \eqref{eq: a-k-approx}, \new{again leading to faster computation}. \resub{Alternative numerical methods such as Goertzel's algorithm \citep{goertzel} may also be used to approximate these integrals.}

From \eqref{eq: fft-solution-II}, we see the utility of using a narrow contour $C$. If $\Delta x$ is small (implying $x_0$ is close to the true solution, since \new{the latter} must be enclosed by $C$), then the second term of \eqref{eq: fft-solution-II} must also be small. For a given desired accuracy in $E_*$ one can thus use fewer grid-points $N$ for the numerical integral \new{in} \eqref{eq: a-k-approx}, obtaining a faster algorithm. This is borne out in practice, since the contour with \resub{$x_0 = \ell +e$} and $\Delta x = \pi/2$ is found to be much more accurate than $x_0 = \pi/2$, \new{$\Delta x = \pi/2-\epsilon$ (the analog to that used in \citealt{2021arXiv210309823S})} for the same $N$.

In a typical setting, one may wish to compute the eccentric anomaly $E_*(\ell,e)$ for many values of $\ell$ (equivalently, time) simultaneously. In this case, computation can be expedited by pre-computing $\sin(-2\pi i k j)$ (which is independent of $\ell$) for each $j=1,...,N-1$. Additionally, the factor $\sin\left(x_0+\Delta x\,e^{2\pi ix}\right)$ appearing in \eqref{eq: ax-def} may be expanded as
\beq\label{eq: double-angle}
    \sin\left(x_0+\Delta x\,e^{2\pi ix}\right) &=& \sin\left(x_0+\Delta x\,\cos 2\pi x\right)\cosh\left(\Delta x\,\sin 2\pi x\right)+i\cos\left(x_0+\Delta x\,\cos 2\pi x\right)\sinh\left(\Delta x\,\sin 2\pi x\right)\\\nonumber
    &=& \left[\sin x_0\cos\left(\Delta x\,\cos 2\pi x\right)^{}+_{}\cos x_0\sin\left(\Delta x\,\cos 2\pi x\right)\right]\cosh\left(\Delta x\,\sin 2\pi x\right)\\\nonumber
    &&\,+\,i\left[\cos x_0 \cos\left(\Delta x\,\cos 2\pi x\right)^{}-_{}\sin x_0\sin\left(\Delta x\,\cos 2\pi x\right)\right]\sinh\left(\Delta x\,\sin 2\pi x\right).
\eeq
Whilst elementary, this is nonetheless useful. With our choice of contour, \new{$x_0 \equiv x_0(\ell,e)$, $\Delta x \equiv \Delta x(e)$, with neither depending on $x$. This implies that} the factors $\sin x_0$, $\cos x_0$ are independent of $x$ (and hence $N$), thus must be computed only once per integral evaluation. Since $\Delta x$ is independent of $\ell$, every other function appearing in \eqref{eq: double-angle} needs to be computed only once per \new{sampling} point, even if a large array is used. This reduces the number of trigonometric function evaluations from $N_\ell \times (N-1)$ to $N_\ell + N-1$ , where $N_\ell$ is the size of the input $\ell$ array, and we note that the points $x = 0$ and $x = 0.5$ are trivial. Once the functions \new{appearing in \eqref{eq: double-angle}} are precomputed, \eqref{eq: a-k-approx} requires only addition and multiplication operations, thus is highly efficient.

\section{Alternative Methods}\label{sec: comparison}

Before presenting the results of our approach, we discuss several popular alternatives. Further details on these can be found in \citet{1999ssd..book.....M} and \citet{1988fcm..book.....D}.

\subsection{Series Solutions}
To obtain an iterative solution to \eqref{eq: Kepler-equation}, we first assume the eccentricity to be small, giving the zeroth-order solution $E_0(\ell,e) = \ell$. Inserting this into the Kepler equation leads to the first-order solution $E_1(\ell,e) = \ell + e\sin \ell$, which may be re-inserted to obtain a second-order solution, \textit{et cetera}. \resub{Taking the infinite limit, this can be written as a Fourier series}
\beq\label{eq: kepler-series-solution}
    E_*(\ell,e) = \ell + 2\sum_{s=1}^\infty \frac{1}{s}J_s(se)\sin s\ell,
\eeq
\citep{1999ssd..book.....M}, where $J_n(x)$ is a Bessel function of the first kind \new{of} order $n$. In practice, we can only compute a finite number of terms; truncating at $s = s_\mathrm{max}$ incurs an error scaling as $e^{s_\mathrm{max}+1}$, which, for small $e$, can be made negligible. As shown in \citet{1970ceme.book.....H}, \resub{the series converges very slowly at large $e$ due to the singularity of Kepler's equation at $e = 1$, $\ell = 0$; the utility of this approach is thus limited in such regimes.}


To implement \eqref{eq: kepler-series-solution}, one must compute the Bessel coefficients for each order $s$ up to $s_\mathrm{max}$. Since these are independent of $\ell$, they may be precomputed for speed \new{(assuming $e$ is fixed)}, implying that the algorithm requires only $N_\ell\times s_\mathrm{max}$ trigonometric function evaluations to compute solutions for a grid of $N_\ell$ mean anomalies. 

\subsection{Numerical Solutions}
A common means of solving monotonic equations such as \eqref{eq: Kepler-equation} is with root-finding, for example via the \resub{quadratic} Newton-Raphson (NR) \new{method} \resub{or the quartic approach of \citet{1988fcm..book.....D}}. For this, one first defines
\beq
    h(E;\ell,e) &=& E - e\sin E- \ell
\eeq
with $h = 0$ at the desired solution. The iterative solutions \resub{are} given by
\beq
    E^\mathrm{NR}_{i+1}(\ell,e) &=& E_i(\ell,e) + \delta_{i1}\\\nonumber
    E^\mathrm{Danby}_{i+1}(\ell,e) &=& E_i(\ell,e) + \delta_{i3}\\\nonumber
\eeq
for $i = 0,1,2,...$, using the definitions
\beq
    \delta_{i1} = -\frac{h_i}{h'_i}, \quad \delta_{i2} = -\frac{h_i}{h_i'+(1/2)\delta_{i1}h_{i}''}, \quad \delta_{i3} &=& -\frac{h_i}{h_i'+(1/2)\delta_{i2}h_i''+(1/6)h_i'''\delta_{i2}^2}
\eeq
for $h_i\equiv h(E_i;\ell,e)$. These require the derivatives $h'(E;\ell,e) = 1-e\cos E$ \textit{et cetera}, 
which are inexpensive to compute given $\sin E$ and $\cos E$. In total, the method requires two trigonometric function evaluations each step, (or $2N_\ell N_\mathrm{step}$ in total across $N_\mathrm{step}$ iterations) as well as a number of (cheap) multiplications and additions.

Both the Newton-Raphson and Danby prescriptions \resub{require also} the initial condition, $E_0(\ell,e)$. Here, we adopt the method of \citet{1988fcm..book.....D}, with
\beq
    E_0(\ell,e) = \ell + 0.85e,
\eeq
but note the additional discussion in \citet{1987CeMec..40..303D}, \resub{\citet{2013CeMDA.115..143C} and \citet{2017CeMDA.129..415E}}. In particular, the \new{first} work notes that convergence to one part in $10^{12}$ can be achieved for all allowed $\ell$ and $e$ values with a maximum of three \new{iterations}. \resub{We note that a similar procedure is given in \citet{1995CeMDA..63..101M}, setting the initial step equal to the solution of a cubic, then performing a single fifth-order root-finding step. Whilst highly accurate, it is not arbitrarily so, and is thus not considered in this work.}

\section{Discussion}\label{sec: discussion}

\begin{figure}
    \centering
    \includegraphics[width=0.45\textwidth]{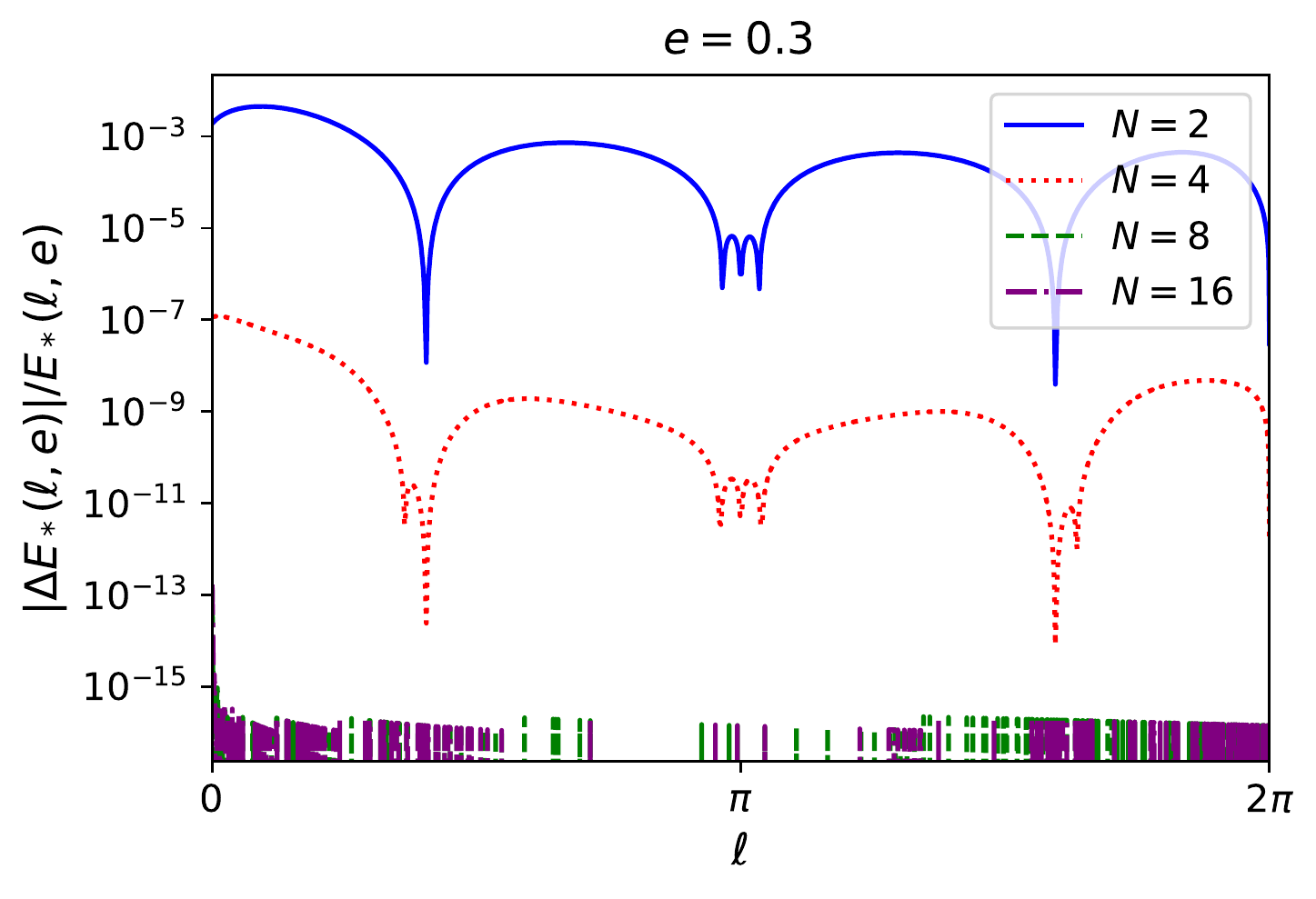}
    \includegraphics[width=0.45\textwidth]{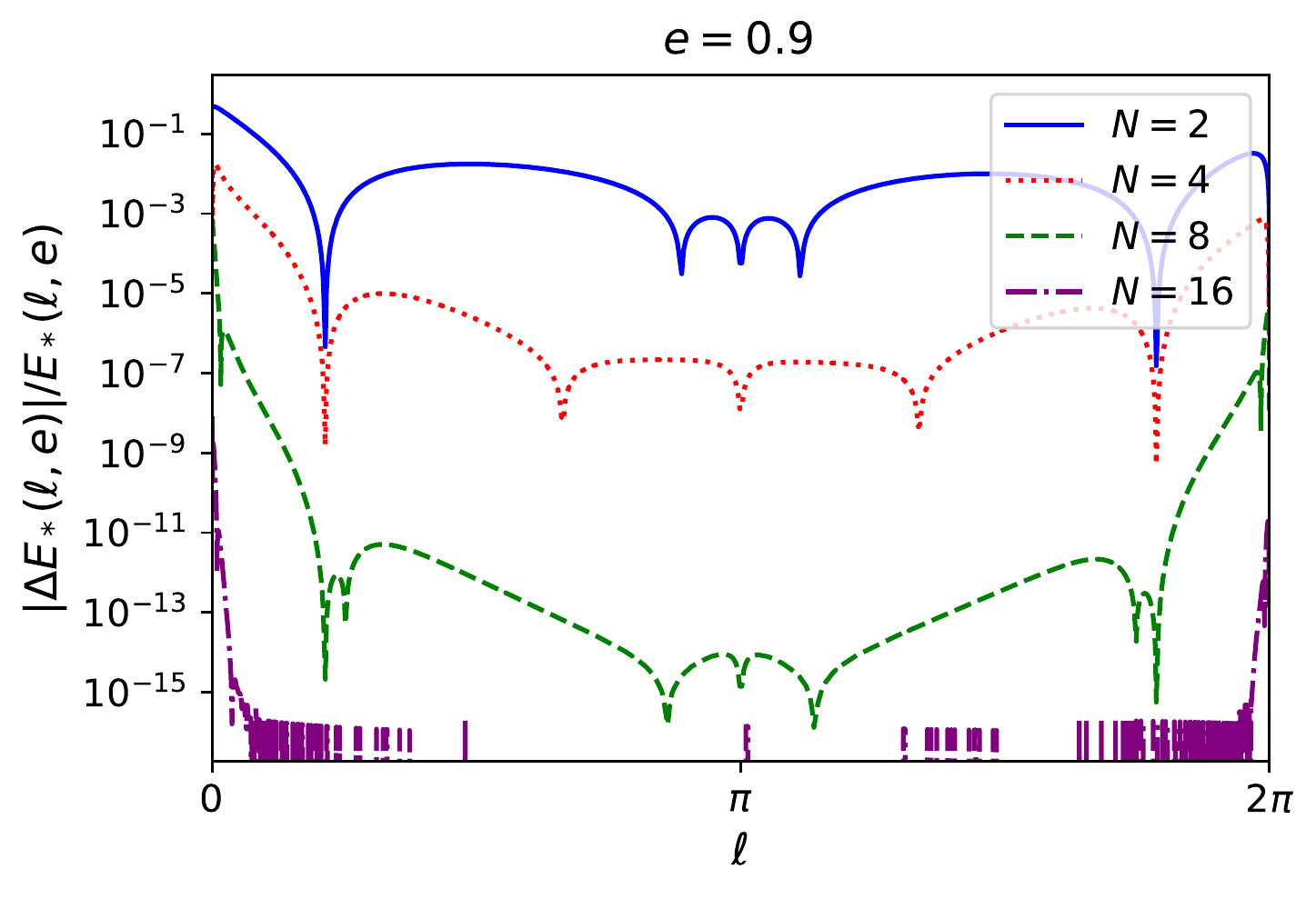}
    \caption{Accuracy of the contour-integration approach to Kepler's equation. Results are shown for two choices of eccentricity $e$, varying the mean anomaly $\ell$ and the number of integration points $N$ and plotting the quantity $\left|\Delta E_*(\ell,e)\right|/E_*(\ell,e)\equiv \left|E_*^\mathrm{true}(\ell,e)-E_*^\mathrm{contour}(\ell,e)\right|/E_*(\ell,e)$. \new{This uses $10^3$ points in $\ell$, generated from an equally-spaced grid in $E$ (giving the true solution).} \resub{The region $\ell>\pi$ can be obtained from $\ell<\pi$ by symmetry, but we include it here for completeness.} As expected, the error in the contour integration decreases rapidly as $N$ increases, and is somewhat larger for greater $e$, due to a wider contour $C$, \new{as in Fig.\,\ref{fig: contour}}. \resub{An additional plot of the error, including its dependence on $e$, is shown in Fig.\,\ref{fig: accuracy2}.} As shown in Tab.\,\ref{tab: time-plots}, our approach \new{requires less computation time relative} to that of conventional methods when the desired precision is held constant.}
    \label{fig: accuracy}
\end{figure}

\begin{figure}
    \centering
    \includegraphics[width=0.9\textwidth]{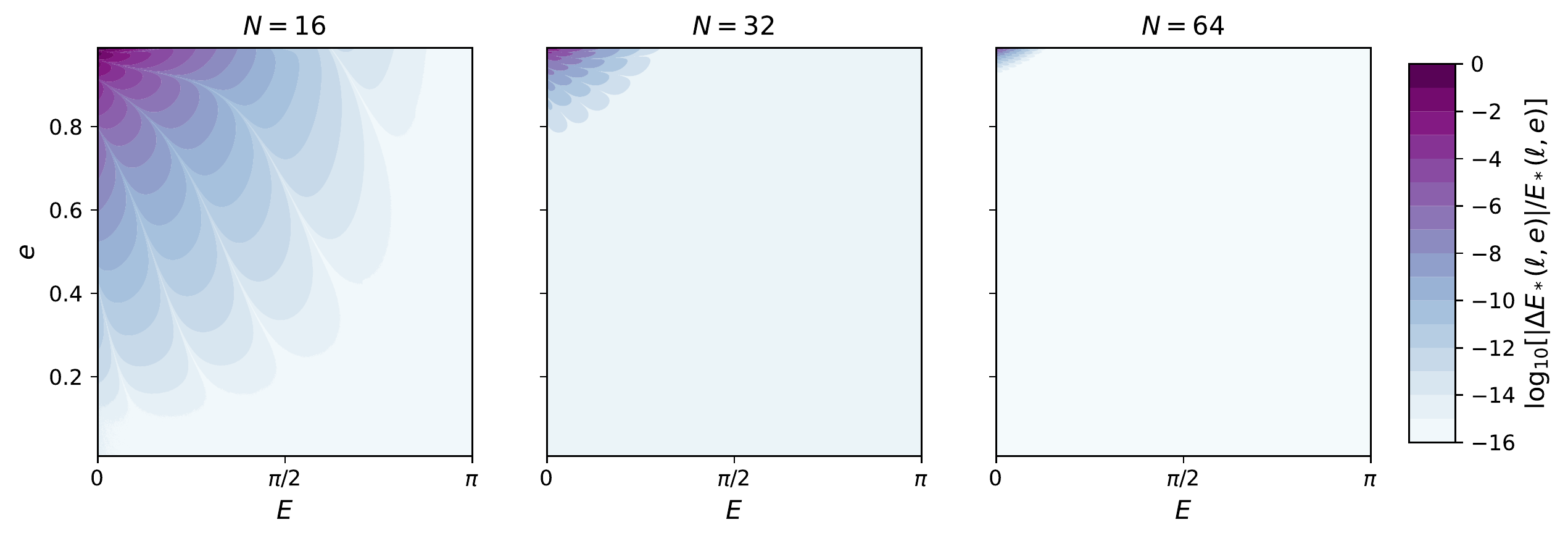}
    \caption{\resub{Fractional error in the contour-integration method as a function of eccentricity $e$ and true eccentric anomaly $E$. We plot the logarithm of the error (defined as in Fig.\,\ref{fig: accuracy}) for three choices of integration step $N$, and show only results for $E<\pi$. $10^3$ points in both $e$ and $E$ (and thus $\ell$) are used to generate this figure. Moving towards the singularity at $e=1$, $\ell=0$, the error increases, but is greatly reduced by choosing a larger $N$.}}
    \label{fig: accuracy2}
\end{figure}

We now compare our prescription (\S\ref{sec: integral-solution}\,\&\,\S\ref{sec: practical}) to the well-known techniques of \S\ref{sec: comparison}. \new{For testing, we generate a grid of $10^6$ equally-spaced points in eccentric anomaly $E$ (defining the `ground truth' solution) and compute the mean anomaly $\ell$ for each via \eqref{eq: Kepler-equation}, assuming $e$ to be fixed.} Figs.\,\ref{fig: accuracy} compares the results from our method \new{(using $10^3$ grid-points)} with the truth for two choices of eccentricity $e$ and various numbers of grid-points, $N$. The results are as expected: the error in the approach falls strongly with increasing $N$,\footnote{The error decreases exponentially with increasing $N$ at fixed $e$ because the error is due to aliasing of higher-frequency Fourier harmonics, and because harmonics of analytic functions fall exponentially at rates proportional to the distance from the integration contour to the nearest singularity.} with accuracy of machine precision achieved for most $\ell$ at $N = 8$ ($N = 16$) for $e = 0.3$ ($e = 0.9$). Furthermore, we find a generally larger error at higher $e$: this is expected since the radius of the integration contour is proportional to $e$, and the solutions lie further from the line $E = \ell$, as in Fig.\,\ref{fig: solutions}. \resub{This is further explored in Fig.\,\ref{fig: accuracy2}, whereupon the error is shown to increase as one moves towards the singularity at $e = 1$, $\ell = 0$. Increasing the number of grid-points greatly reduces this however, highlighting the importance of choosing $N$ appropriately.}

To place our method in context, we compare it to the three alternative prescriptions (series, Newton-Raphson, and Danby) discussed in \S\ref{sec: comparison}. For this purpose, we implement each in \new{\textsc{C++}}, paying close attention to efficiency.\footnote{\new{\textsc{C++}} code implementing all of these can be found at \href{https://github.com/oliverphilcox/Keplers-Goat-Herd}{github.com/oliverphilcox/Keplers-Goat-Herd}.} To ensure each method has a similar level of accuracy, we gradually increase the \new{precision} of each method (either by increasing the number of grid-points in the numerical integral, or the number of iterations) until a mean absolute precision of $10^{-12}$ is obtained relative to the `ground-truth' solution. 

\begin{table}
    \centering
    \begin{tabular}{l | cc | cc | cc |}
    \multirow{2}{*}{Method} & \multicolumn{2}{c}{$e = 0.1$} & \multicolumn{2}{c}{$e = 0.5$}  & \multicolumn{2}{c}{$e = 0.9$} \\
     & $N_\mathrm{it}$ & Time & $N_\mathrm{it}$ & Time & $N_\mathrm{it}$ & Time\\
    \cmidrule{1-1}
    \cmidrule(lr){2-3}
    \cmidrule(lr){4-5}
    \cmidrule(lr){6-7}
    Newton-Raphson & 3 & 97.5 & 4 & 133 & 5 & 192\\
    Danby (1988) & 2 & 82.8 & 2 & 82.8 & 3 & 127\\
    Series & 11 & 116.2 & 47 & 516 & - & -\\
    \textbf{This Work} & \textbf{5} & \textbf{35.1} & \textbf{7} & \textbf{41.1} & \textbf{18} & \textbf{65.9}\\
    \end{tabular}
    \caption{Computation time (in milliseconds) required to solve Kepler's equation \eqref{eq: Kepler-equation} for $10^6$ $\ell$ \new{points (equally spaced in $E$)} using three popular methods and that of this work. The Newton-Raphson and \citet{1988fcm..book.....D} \resub{methods} are quadratic (quartic) root-finders, whilst the series approach is an expansion in eccentricity. 
    The number of iterations (or integration points) $N_\mathrm{it}$ was chosen by repeating the calculation until a mean absolute error below $10^{-12}$ \new{(relative to the true solution)} was obtained for the sample. All algorithms were implemented in \new{\textsc{C++}} and run on a 2019 MacBook Pro with a 6-core 2.6\,GHz Intel i7 processor. The series \resub{was very slow to converge} for
    $e = 0.9$, hence no timings are shown. We find our approach to be significantly faster than \new{the alternatives considered.}}
    \label{tab: time-plots}
\end{table}

The final hyperparameter values and runtimes for the $N_\ell = 10^6$ array of mean anomalies are shown in Tab.\,\ref{tab: time-plots} for three values of the eccentricity. As expected, both the Newton-Raphson and Danby methods converge quickly, with the latter requiring a maximum of three iterations even at the most extreme case of $e = 0.9$. The Danby prescription requires more computational operations per iteration, thus its runtime is only slightly reduced \new{relative to Newton-Raphson}, despite the fact that the number of iterations is roughly halved. Significantly more iterations are required for the series solution of \eqref{eq: kepler-series-solution}, particularly as $e$ increases. Given its nature as a perturbative solution, this is unsurprising, and we conclude that it is of most use only for orbits of very low eccentricity. Finally, we consider our approach. In all cases, we require \new{$N\leq 18$} grid-points in the numerical integral (with $N$ rising as $e$ increases, as in Fig.\,\ref{fig: contour}), and find that the computation time is significantly reduced (by a factor $\sim$~$2-3$) compared to any of the alternative prescriptions. This may be understood by considering the number of trigonometric operations involved: assuming $N_\ell\gg N$, the total is proportional to $N_\ell$, and thus independent of the number of steps. Each integration step simply requires multiplication and addition operations, giving a faster speed.

To summarize, this work demonstrates that new methods to solve the Kepler equation can provide solutions outperforming those of established methods over a wide range of parameter space. Our approach is based on contour integration, using methods developed to solve the `geometric goat problem' in \citet{Ullisch} and recently applied to the dynamics of spherical collapse in \citet{2021arXiv210309823S}. We expect such approaches to be similarly applicable to other problems that require the solution of transcendental equations.

\section*{Data Availability}
The data underlying this article will be shared on reasonable request to the corresponding author. \textsc{C++} \resub{and \textsc{python}} implementations of our code are available at \href{https://github.com/oliverphilcox/Keplers-Goat-Herd}{github.com/oliverphilcox/Keplers-Goat-Herd}.

\section*{Acknowledgments}
\resub{We thank Tom Dickens, Hanno Rein, Ian Weaver and the Twitterverse for insightful feedback. We are additionally grateful to the referee, Manuel Calvo, for an insightful report.} OP acknowledges funding from the WFIRST program through NNG26PJ30C and NNN12AA01C. 

\bibliographystyle{mnras}
\bibliography{bib}





\bsp	
\label{lastpage}
\end{document}